\documentclass[%
	aip,
	cha,%
	amsmath,amssymb,
	reprint,%
   ]{revtex4-1}
   \usepackage{color}
   \usepackage{graphicx}
   \usepackage{dcolumn}
   \usepackage{bm}
   \usepackage{amsfonts}
   \usepackage{sidecap}
   \usepackage{float}
   \usepackage{parskip}
   \usepackage{grffile}
   
   \usepackage{color}
   \usepackage{hyperref}
   \usepackage{hhline}
   \usepackage{mathtools}
   \usepackage[normalem]{ulem}
   \makeatletter
   \def\@eqnnum{{\normalsize \normalcolor (\theequation)}}
	\makeatother
\begin{document}

\title{Engineering chimera patterns in networks using heterogeneous delays}
\author{Saptarshi Ghosh}
\affiliation{Complex Systems Lab, Discipline of Physics, Indian Institute of Technology Indore, Khandwa Road, Simrol, Indore 453552}
\author{Sarika Jalan}
\email{sarikajalan9@gmail.com}
\affiliation{Complex Systems Lab, Discipline of Physics, Indian Institute of Technology Indore, Khandwa Road, Simrol, Indore 453552}
\affiliation{Discipline of Bio-Science and Bio-Medical Engineering, Indian Institute of Technology Indore, Khandwa Road, Simrol, Indore-453552}
\date{\today}

\begin{abstract}
Symmetry breaking spatial patterns, referred as chimera state, have recently been catapulted into the limelight due to their coexisting coherent and incoherent hybrid dynamics. Here, we present a method to engineer a chimera state by using an appropriate distribution of heterogeneous time delays on the edges of a network. The time delays in interactions, intrinsic to natural or artificial complex systems, are known to induce various modifications in spatiotemporal behaviors of coupled dynamics on networks. Using a coupled chaotic map with the identical coupling environment, we demonstrate that control over the spatial location of the incoherent region of a chimera state in a network can be achieved by appropriately introducing time delays. This method allows engineering tailor-made one cluster or multi-cluster chimera patterns. Furthermore, borrowing a measure of eigenvector localization from spectral graph theory, we introduce a spatial inverse participation ratio which provides a robust way for identification of the chimera state. This report highlights the necessity to consider the heterogeneous time delays to develop applications for the chimera states in particular and understand coupled dynamical systems in general.
\end{abstract}
\maketitle

\begin{quotation}
The study of time-delayed dynamics on a complex network is a field of high interest for both its fundamental significance in the study of non-linear systems and its applicability in understanding various networks modeled after biological, chemical and artificial or natural systems which intrinsically possess the information propagation delays among their constituent entities. Both homogeneous and heterogeneous propagation delays
between the nodes can induce a significant change in the dynamical properties of a complex system represented
by networks. Specifically, a massive amount of work has been done to investigate the impact of delays on
various types of synchronization. Among them, partially synchronized state of chimera received particular attention due to the coexistence of coherent and incoherent dynamics. Since its discovery, the hybrid spatial patterns of chimera state have become a prime branch of complex systems research primarily due to fascinating
phenomena of symmetry breaking in identical systems along with many other applications
ranging from getting insight into the transition between coherence to incoherence and neuroscience.
Despite numerous studies on chimeras, controlling chimera state has remained a tricky problem due to the peculiar nature of chimera patterns. Here, we propose a scheme to engineer the chimera patterns by
suitable placement of delayed edges in a network. Furthermore, inspired by the eigenvector
localization measures, we suggest a measure to identify chimera states.
\end{quotation}

\section{Introduction}
Network science has attracted an avalanche of investigations for understanding underlying interaction
patterns of natural and artificial complex systems due to its wide applicability in various fields
ranging from sociology to economics and in engineering problems ~\cite{network_rev1,network_rev2,network_rev3}. Among the current
advancements in the study of the emergent collective behavior of a network, chimera state deserves special
attention due to the emergence of coexisting coherence and incoherence dynamics led by symmetry breaking
in a network of identical elements~\cite{chim_rev1}. Since landmark works of Kuramoto {\it et al.}~\cite{chim_def1} and Abrams {\it et al.}~\cite{chim_def2} for phase oscillators, chimera state has been reported for numerous time-discrete as well as time-continuos dynamical systems~\cite{chim.discrete.cont,chim_rev1}. Recently, the chimera state has been extended to multiplex networks~\cite{sapta_mul_chim}.
Although initially non-local couplings as well as special initial states~\cite{chim_def1,chim_def2} have been reported to be the necessary condition for the emergence of chimera, subsequent studies have found chimera states for purely local~\cite{chim.local_coup}, completely global~\cite{chim.global} as well as for random initial conditions~\cite{Random_init_chim}. A plethora of experimental as well as theoretical studies~\cite{chim_rev2} has been carried out to put forward a better understanding of its illusive nature~\cite{chim_rev3}. 

Various biological processes such as uni-hemispheric sleep in mammals are known to show chimera-like states~\cite{uni_sleep}. Chimera patterns are reported to emerge for neural activities related to several brain diseases ~\cite{brain_disease}. Recently, {\it Andrzejak et al.} presented probable relations between identifying and diagnosing of the epileptic seizure and chimera state~\cite{EEG_chim}. Furthermore, inhibition is known to play a pivotal role in regulating high synchronization among neurons which leads to
a destruction of complex pathways in the brain and eventually yielding diseases like the epileptic seizure~\cite{inhi}. Recent works have established a bridge between the emergence of chimera state and the inhibition present in the system~\cite{chim_inhi} which further highlights the importance of chimera states in neural networks. 

Due to its potential applicability as well as fundamental significance, there have been persistent efforts to control the chimera states~\cite{chim_control}. There have been several studies on control of the parameter regime as well as regions of (in)coherence for the emergence of chimera states~\cite{sapta_mul_chim_control}. In earlier works~\cite{chim_delay,chim_delay2}, it has been shown that one of the approaches to control chimera is to introduce delayed interaction between the nodes of the network. Here, in this report, we approach the problem of managing the chimera states via introducing heterogeneous delay on the edges of a network. The presence of heterogeneous delays in a network is more realistic in the context of real-world networks where interactions (edges) between the pairs of nodes are subjected to heterogeneous perturbations from its surroundings. 
Due to these physical constraints, the heterogeneous delayed interactions between nodes is a naturally occurring phenomenon in various complex systems represented by networks~\cite{delay_book}. Delay has been shown to be responsible for a plethora of novel emerging phenomena in different dynamical systems~\cite{delay_papers}. Here, we demonstrate that the chimera patterns i.e., the emergence of the region(s) of incoherence can be controlled by suitably introducing the heterogeneous delays (and thus breaking the dynamical symmetry) in a particular portion of the network. More specifically, our scheme allows generating tailor-made chimera patterns which can be accurately tuned to the placement of delayed nodes. Besides, we introduce an entirely new way to detect the chimera by borrowing an eigenvector localization measure from spectral graph theory.

This report is arranged as follows: Sec II includes a brief discussion on the coupled dynamics on the
network followed by the introduction of the new measure to detect the chimera states. Sec III presents
results on the production of engineered chimera states in regular networks
by appropriate introduction of the heterogeneous delays. Sec IV presents a detailed study on the impact of heterogeneity present in a delayed network. To the end, we include a section summarizing the results and their
importance in the context of understanding the chimera states.

\section{Theoretical model.}

\subsection{Dynamical evolution on networks.} In this report, we use a network consisting of $N$ nodes interacting via $N_c$ edges. The dynamical state of the nodes at time $t$ can be represented by a real variable $x_i(t)\in \mathbb{R}, \forall i=1,...,N$. The time evolution of the dynamical state of nodes can be written in terms of a time discrete map $x_i(t+1)=f(x_i(t))$ where we consider famous logistic map $f(x)=\mu x (1-x)$ in chaotic regime ($\mu=4.0$)~\cite{logs_chaos} as local dynamics. The simplistic framework of logistic map have been used to understand diverse spatio-temporal phenomena in a wide range of real world networks~\cite{logs} among which chimera has also been shown in both single~\cite{chim.discrete.cont} and multiplex~\cite{sapta_mul_chim} networks. Adding the network architecture, the dynamical evolution equation for the whole network can be written as \cite{sapta_mul_chim}

\begin{equation}
x_i(t+1)=f(x_i(t))+\frac{\varepsilon}{(k_i)} \sum_{j=1}^{N} A_{ij}[ f(x_j(t))-f(x_i(t)) ]
\label{eq.evol}
\end{equation}
where $A_{ij}$ represent the element of the adjacency matrix $A$ and takes value 1 if $i^{th}$ and $j^{th}$ nodes are connected and $0$ otherwise. Further, $k_{i}$ = $\sum_{j=1}^{N}A_{ij}$ depicts degree of the $i^{th}$ node and $\varepsilon$ represents the overall coupling constant ($0\le \varepsilon \le1$). We have chosen a non-locally coupled regular ring network (Fig.~\ref{fig3}) to showcase our findings. 

We introduce delays in the edges of the network which are represented by a delay matrix $D$ consisting of elements $D_{ij} \ni D_{ij}=\tau$ if edge $A_{ij}$ is delayed and $0$ otherwise. The entry $\tau$ takes
the delay value. The entries of the delay matrix depend on the intended investigation. For example, 
$\tau$ takes the same value for all the edges of the network in the case of homogeneous delay. Whereas $\tau$ is a random variable taken from a uniform distribution bounded by $0\le\tau\le\tau_{max}$ for heterogeneous delay case where delayed nodes are placed in particular spatial positions of the network. Thus, $\tau_{max}$ represents the upper limit of the random entries of the delay matrix. 
Note that a delayed node means the node with all the edges originating from it are delayed. The dynamical evolution (Eq.\ref{eq.evol}) can be rewritten to incorporate delay as \cite{het_delay}
\begin{equation}
x_i(t+1)=f(x_i(t))+\frac{\varepsilon}{(k_i)} \sum_{j=1}^{N} A_{ij}[ f(x_j(t - D_{ij}))-f(x_i(t)) ]
\label{eq.devol}
\end{equation}
$D_{ij}$ is the time taken for the dynamical information to reach from the node $i$ to the node $j$ and vice-versa. Here, we consider a symmetric delay matrix.

\begin{figure*}[t]
 \centerline{\includegraphics[width=4.5in,height=3.0in]{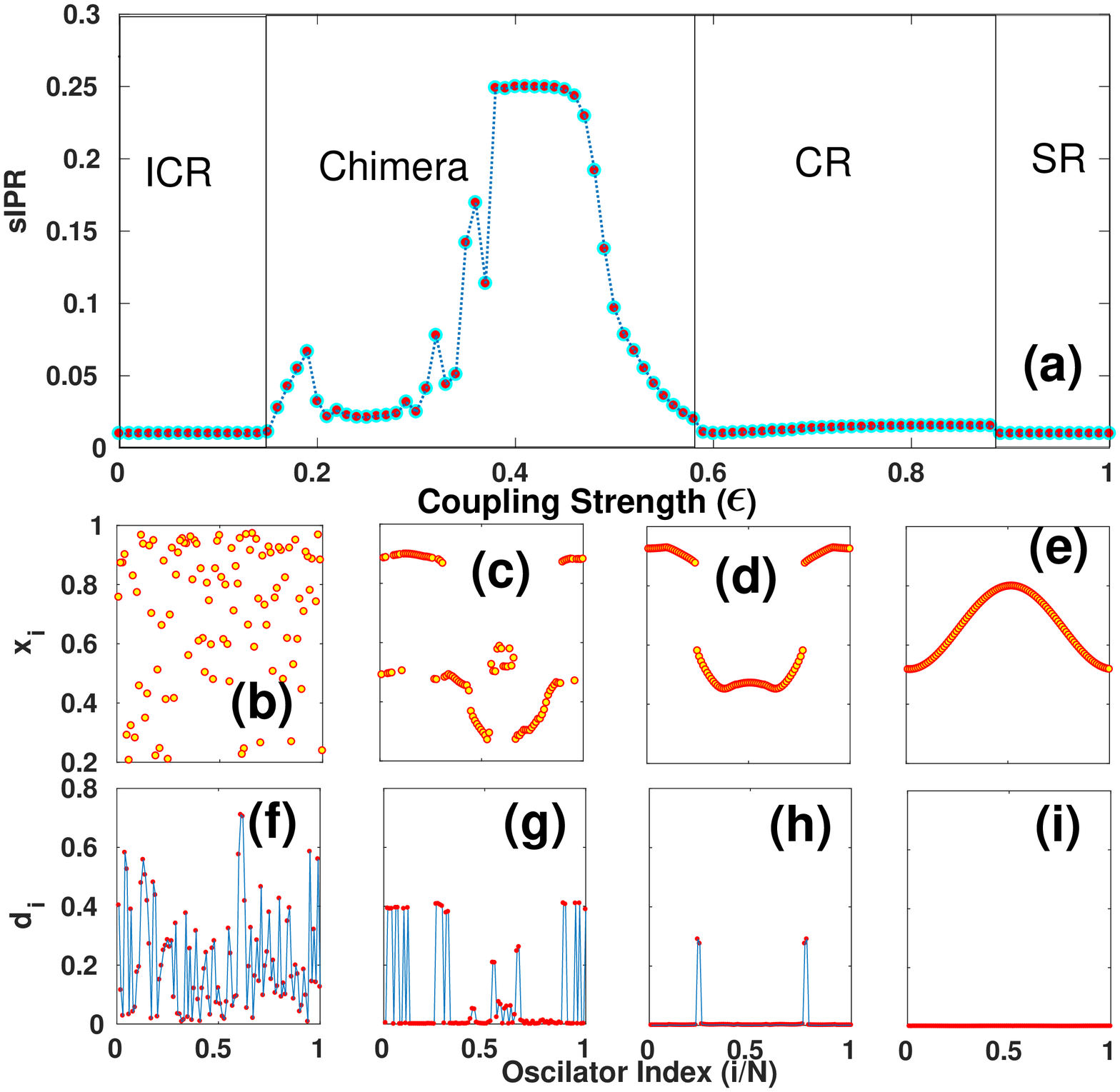}}
 \caption{(Color online) (a) sIPR profile as a function of coupling strength for a undelayed regular network depicting transition from the incoherent (ICR) to the coherent state (CR) via a chimera state. 
The exact synchronous region (SR) produces a undefined sIPR and thus is forced to take ($1/N$) value. 
Diagram (b) - (i) represent, respectively, snapshots and Laplacian profiles of the
dynamical states for the regular network ($S^1$ ring) corresponding to
various points in the sIPR profile plotted in (a). (b)\&(f) $\varepsilon=0.1$ and $sIPR=0.029$, 
(c)\&(g) $\varepsilon=0.34$ and $sIPR=0.048$, (d)\&(h) $\varepsilon=0.4$ and $sIPR=0.25$, (e)\&(i) $\varepsilon=0.76$ with $sIPR=0.015$. Other Parameters are network size$(N)$ $=100$, node degree $(k)$ $= 64$.}
 \label{fig2}
 \end{figure*}

\subsection{Chimera state.} A chimera state is defined as a hybrid dynamical state consisting of coexisting coherent and incoherent domains which appear in structurally symmetric networks. As mentioned, we consider a regular network architecture with the periodic boundary condition ($S^1$; ring) to showcase the occurrences of the chimera state. The coherent dynamical sate on the network can be depicted as \cite{chim.discrete.cont}
\begin{equation}
\lim\limits_{N \rightarrow \infty} \lim\limits_{t \rightarrow \infty}\sup\limits_{i,j  \in U_{\xi}^N (x)} \mid{x_i(t)-x_j(t)} \mid \rightarrow 0 \, \, \text{for} \, \, \xi\; \rightarrow 0
\label{eq.cohr}
\end{equation}
where $U_{\xi}^{N} (x) = \{ j : 0 \le j \le N, \mid{\frac{j}{N} - x} \mid <  \xi \}$ depicts the spatial neighborhood of any point $x \in S^1$, i.e., circle structure of the 1D ring network. Geometrically, a 
smooth spatial profile of the dynamical variable $x_i(t)$ in $x_i - i$ plane (in the continuum limit $N \to \infty$ ) depicts a coherent state. A profile having  domains of the smooth spatial curve broken by the discontinuous regions represents a chimera state.

\subsection{Spatial Inverse participation Ratio (sIPR): A measure for identification of chimera.} Due to the
peculiarity of the spatial profiles of chimera state, a plethora of measures had been put forward in literature \cite{chim_measure,chim_delay}. In this article, we propose a new measure borrowed from the eigenvector localization concepts \cite{loc_prio} of the spectral graph theory. The inverse participation ratio (IPR), in the classical sense, depicts the contribution of elements in a eigenvector(state) \cite{ipr_def}. Following the trend, we define the spatial inverse participation ratio (sIPR) as

\begin{equation}
sIPR=\frac{\sum_i (\langle d_i  \rangle_t)^4}{\{\sum_i (\langle d_i\rangle_t)^2\}^2}
\label{sipr}
\end{equation}
where $d_i =|(x_{i+1}(t)-x_{i}(t))-(x_{i}(t)-x_{i-1}(t))|$. $\langle d_i \rangle_t$ depicts an average value of $d_i$ over time. Overall $d_i$ depicts a discrete second-order differentiation (Laplacian in general) representing the relative spatial distances between neighboring nodes.
A high value of $d_i$ corresponds to a large spatial gap between the neighbors of the $i^{th}$ node, appearing as a discontinuity in the spatial profile (in $x_i -i$ plane), whereas a low value of $d_i$ indicates that the $i^{th}$ node is spatially close to its neighboring nodes. If all the $d_i$ take high values (Fig.~\ref{fig2}(f)), It represents the incoherent state where all neighboring nodes have large spatial distance between them (Fig.~\ref{fig2}(b)). If all the $d_i$ take low value (Fig.~\ref{fig2}(i)), it corresponds to the coherent state where all the neighboring nodes are close by ((Fig.~\ref{fig2}(e))) and form a smooth spatial profile.  
However, for both the cases, the participation of the elements are similar, i.e., all of the entries of $d_i$ for different nodes are either high valued or low valued. This data trend of similar entries results in a low value of sIPR as dictated by the definition of the traditional IPR~\cite{loc_prio}. 
However, for hybrid patterns of chimera state, nodes forming the CR takes low $d_i$ values whereas the nodes in ICR produce high $d_i$ values (Fig.~\ref{fig2}(c),(g) \& (d),(h)). This breaks the data trend of similar values in $d_i \, ;\forall i$ and we demonstrate that this break in the trend can be picked up by the sIPR value and can be used to identify the chimera states (Fig.~\ref{fig2}). Therefore, a high value of sIPR shows the presence of both high and low $d_i$ entries denoting coexistence of CR and ICR forming a chimera state. On the other hand, a low value of sIPR denote all entries of $d_i$ are either high or low representing an incoherent and a coherent state, respectively. Fig.~\ref{fig2}(a) demonstrates a sIPR profile for an undelayed regular ring network as a function of the coupling strength where a transition from the incoherent to the coherent state via chimera is clearly visible.
Traditionally, the IPR  value of a eigenvector for a network of dimension $N$ is bound by $\frac{1}{N} \le IPR \le 1$ \cite{loc_prio}. We demonstrate that the incoherent or coherent states produce sIPR values close to $\frac{1}{N}$ (Fig.~\ref{fig2}(a)), Whereas the chimera state produces a significantly higher value of sIPR. Due to the lower bound of the IPR, the sIPR automatically assumes the value $\frac{1}{N}$ (with N being network size) for a non-chimera state without needing any threshold value unlike other measures of chimera state~\cite{chim_measure}. A point to note here that, due to the low values of $d_i$ for the coherent states (Fig.~\ref{fig2}(i)), sIPR value may become undefined. In that case, we manually set the IPR value as $\frac{1}{N}$ to maintain the similarity (Fig.~\ref{fig2}(a), SR region). To summarize, sIPR captures the similarity or dissimilarity in the values of $d_i$ to identify chimera state such that, it yields values closer to  $\frac{1}{N}$ for the coherent or incoherent case, whereas it produces a high value for the chimera state.

\begin{figure}[t]
 \centerline{\includegraphics[width=2.8in,height=1.5in]{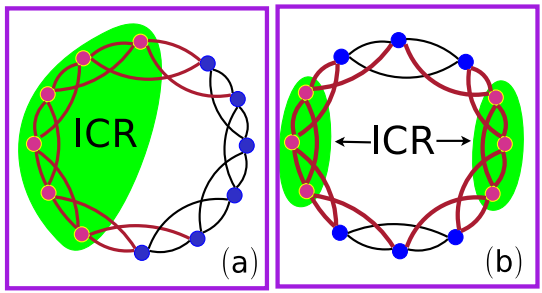}}
 \caption{(Color online) Schematic diagram depicting 
various heterogeneous delays configurations for a regular ring network which we have used
to design chimera states. 
The shaded areas denote the nodes experiencing heterogeneous delay. 
(a) A single cluster of delayed nodes (those in the shaded region) 
producing corresponding ICR region (corresponds to Fig.~\ref{fig4}(b)).
(b) Two clusters of delayed nodes (those in the two different shaded regions) 
separated by the undelayed nodes producing multiple ICRs separated by CR 
(corresponds to Fig.~\ref{fig4}(h)). 
The schematic depiction of regular network is for node degree$(k)$ $= 4$,
however, for the actual simulations the parameters are described in Fig.~\ref{fig2}.}
 \label{fig3}
 \end{figure}

\section{Results.}

\subsection{Engineered chimera state with heterogeneous delay.} 
\begin{figure}[b]
	\centerline{\includegraphics[width=3.9in,height=4.0in]{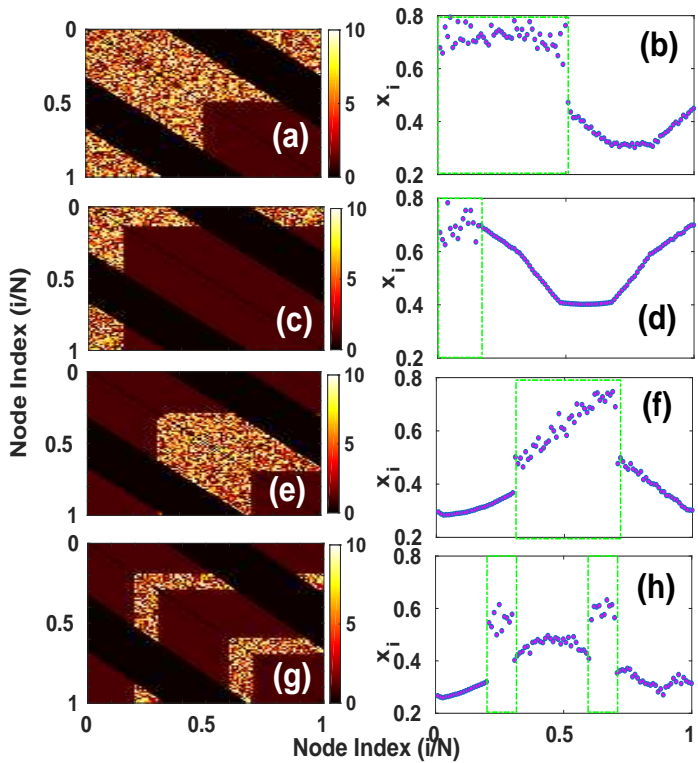}}
	\caption{(Color online) Delay matrices and corresponding snapshots of chimera state engineered by heterogeneously distributed delays. The delay matrices are overlapped with the adjacency matrices to showcase both the delayed and the undelayed edges. (a) \& (c) Regular network with a large and relatively smaller cluster of the delayed nodes, respectively. These delay configurations result in a large ICR denoted in (b) and a smaller ICR in (d). (e) \& (f) Delay configuration and corresponding snapshot of a chimera state, respectively with different location of the ICR. Note due to the $S^1$ symmetry of the regular ring lattice, the positions are not unique. (b), (d), \& (f) correspond to schematic diagram Fig.~\ref{fig3}(a), and 
(g) and (h) represent two delayed clusters resulting in multi-chimera state, corresponding to 
the schematic  diagram Fig.~\ref{fig3}(b)). The (green) boxes represent clusters of the 
delayed nodes introduced in the network. Other parameters are $\varepsilon=0.77$, $\tau_{max}=10$ and rest are same as Fig.~\ref{fig2}}
	\label{fig4}
	\end{figure}

Information transfer between a pair of interacting nodes takes a finite propagation time to reach from one node to another node. Therefore, a delayed interaction, particularly heterogeneous delay is an intrinsic property of several natural and man-made networks. For example, in an aircraft network, travel time between two airports is subjected to weather conditions leading to heterogeneous delays in the system. Similarly, in a bio-chemical PPI (Protein–Protein Interaction) network, the interaction time between two proteins is subjected to its chemical environment. Here, we investigate the impact of heterogeneously distributed delay in a network on occurrence of chimera states.
A chimera state consists of a coherent region (CR) and an incoherent region (ICR) which coexist simultaneously. There are several investigations on controlling the positions of the CR/ICR, to produce custom-made chimera patterns \cite{chim_control}. We approach this problem of controlling chimera by introducing delayed nodes in the network. The delayed node means all the edges originating from that node has a delay selected randomly between $0$ to $\tau_{max}$. We choose a value $\tau_{max}$ such that it is larger than the intrinsic time scale of the underlying dynamical system which is unity in the case of time -discrete logistic map considered here~\cite{tau}. Here we show that the position of the ICR can be controlled by suitably placing the delayed nodes in a preferred spatial location. Furthermore, this scheme does not depend on the choice of the $\tau_{max}$ (See Sec.~\ref{SM}). A schematic diagram depicting the protocol for delay distribution on the nodes of a regular ring network is presented in Fig.~\ref{fig3}.
Fig.~\ref{fig3} depicts the clusters containing the delayed nodes
(denoted by red circles in the shaded region) contributing to the
ICR (corresponding to the spatial profiles depicted in Fig.~\ref{fig4} (b) \& (h)).
Fig.~\ref{fig3}(a) presents the case where a cluster consisting
of neighboring nodes are heterogeneously delayed. This design
produces a chimera state with one ICR and one CR (Fig.~\ref{fig4} (b)).
Whereas, Fig.~\ref{fig3}(b) presents the case of two clusters of delayed nodes producing multiple ICRs separated by a CR (Fig.~\ref{fig4}(h)).

Fig~\ref{fig4} presents different types of chimera states with a corresponding delay matrix profile which is engineered based on the our desired output. Fig~\ref{fig4}(a) depicts a color profile of the delay matrix. 
The heterogeneous delays are introduced in half of the nodes which are clustered together in the
terminal position. This protocol results in a chimera state with one CR and ICR as 
depicted in Fig~\ref{fig4}(b). Note that exact position of the ICR coincides with that of 
the delayed nodes (green boxes in Fig.~\ref{fig4}). To highlight the effect, we consider a 
similar delay matrix as for the previous example, however, with less number of the 
delayed nodes (Fig~\ref{fig4}(c)). As expected, the spatial profile of the chimera state contains 
a reduced ICR at the position coinciding with the position of the delayed nodes. 
These dependence of the ICR on the position of delayed nodes hold good even if we introduce delay in the 
central part of the spatial profile as depicted in Fig~\ref{fig4}(e) and (f) which results in one ICR 
bounded by two CRs. We further demonstrate that by appropriate engineering of the delay matrix
we  can  produce multi-chimera states with 
multiple ICRs. Fig~\ref{fig4}(g) depicts that the delays are introduced in the terminal positions 
separated by a region of undelayed nodes. This brings forward a multi-chimera state with two ICRs separated by a CR (Fig~\ref{fig4}(h).

Therefore, the location of the ICR(s) can be controlled by adopting appropriate protocol
of distribution of delays on the edges of the network. Note that due to $S^1$ symmetry of the regular 
ring network considered in this paper, there is no unique position of the nodes. However, we have 
referred to the unique numerical naming of the nodes (node number 1 to node number N) to refer their 
positions for an easy depictions of our results. The relative positions of the single-cluster or 
multi-cluster chimera state reflect that an appropriate distribution of
heterogeneous delay can accurately engineer the 
spatial profile of the chimera states regardless of the nomenclature of the nodes.

\begin{figure}[t]
	\centerline{\includegraphics[width=4.0in,height=1.5in]{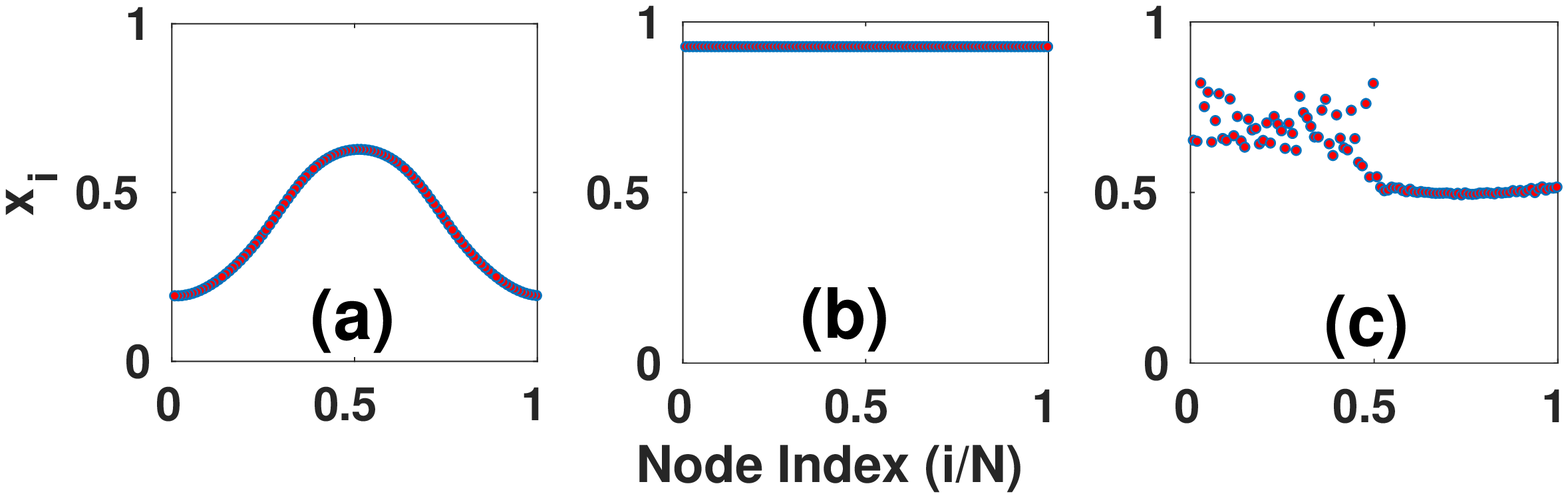}}
	\caption{(Color online) Snapshots of the dynamical state of the regular network for different delay configurations with (a) being no delay configuration, (b) being homogeneous delay ($\tau=1$) case, (c) being heterogeneous distributed delays case ($\tau_{max}=10$). Note that the typical spatial profile for chimera state is visible for (c). Other Parameters are $\varepsilon=0.61$ and rest are same as Fig.~\ref{fig2}}
   \label{fig5}
   \end{figure}

\subsection{Impact of heterogeneous delay on the emergence of chimera states} 
The previous section demonstrates that the chimera patterns can be engineered by a suitable
placement of the heterogeneous delays. However, this fine control is possible only in the high coupling regions.
In these regions,
the chimera state does not appear for the undelayed or homogeneously delayed case. 
Furthermore, a protocol of 
distribution of heterogeneous delays on all the nodes of a network, i.e. all the nodes in the network are delayed, also does not produce the chimera states. For this protocol, a direct transition from the 
incoherent to the coherent states takes place. In the following, we present an elaborate discussion of this point.

\begin{figure}[b]
	\centerline{\includegraphics[width=3.8in,height=3.0in]{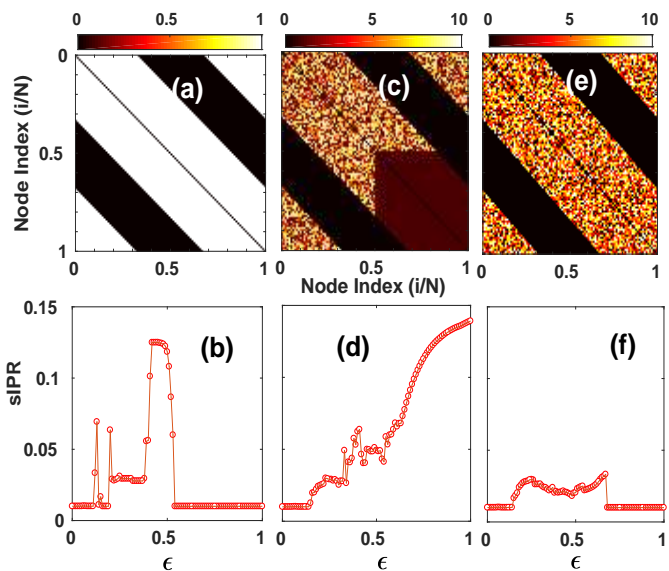}}
	\caption{(Color online) Delay matrices representing the heterogeneous delay induced in the network and corresponding sIPR profile for the delay configuration. the delay matrices are overlapped with the adjacency matrices to show both the delayed and the undelayed edges. (a) A network with homogeneous delay ($\tau=1$) with a typical color profile for 1D ring lattice adjacency matrix (b) sIPR profile indicating incoherent, chimera and coherent states respectively. (c) Delay matrix with partial heterogeneous delay represented by the mosaic pattern extended by shaded region representing edges with no delay. (d) sIPR profile indicating incoherent and subsequent chimera state (e) Delay matrix with full heterogeneous delay represented by mosaic pattern inhibiting whole adjacency matrix (f) sIPR profile with direct incoherent to coherent transition. Other parameters are $\tau_{max}=10$ (for c \& e) , and  rest are same as Fig.~\ref{fig2}.}
	\label{fig6}
	\end{figure}

For the partial heterogeneous delay case (i.e., only few nodes are delayed), an incoherent evolution 
is observed in the weak coupling region, followed by a chimera state in the mid coupling range. 
This chimera state appearing in the mid-coupling region is completely random and cannot be controlled using appropriate placement of the delays. However, the high coupling range yield a drastic change
in the dynamical evolution and we achieve a direct relation between the position of delayed nodes and
ICRs.
Note that, for both the protocols of the undelayed and homogeneously delayed networks, 
the high coupling regions yield a coherent dynamics \cite{sapta_mul_chim} as depicted 
in Fig~\ref{fig5}(a) and (b). Fig~\ref{fig5}(c) demonstrates a chimera state in the same region 
engineered by suitably placed heterogeneous delays. Therefore, we can conclude
that the heterogeneous delays not only can lead to an enhancement in the parameter region for which chimera states appear but also offer a control in a limited parameter regime where we can produce tailor-made chimera patterns.

Furthermore, we find that a complete envelopment of edges by heterogeneous delays can be harmful for the chimera states. In the previous section, we demonstrated that the ICR coincides with the heterogeneous delayed nodes. Using this approach, we had shown that the production of both single and multi-cluster chimera 
states can be achieved. However, we had found that if heterogeneously distributed delays span over all the edges in the network, the chimera state is ceased to exist. Fig~\ref{fig6}(a) presents a 
typical sIPR profile for a homogeneous delay case. We observe a transition from the incoherent to  the coherent state via a chimera state. The measure sIPR cannot clearly distinguish between the completely 
incoherent and a completely coherent state. For example,  Fig.~\ref{fig5}(b) presents that
homogeneous delays (corresponding delay matrix is depicted in Fig.~\ref{fig6}(a))
renders the coherent state for the high coupling region. Fig.~\ref{fig6}(b) presents the 
enhanced parameter regime for the appearance of chimera states in the 
partial heterogeneous delay case where we can also observe chimera state in the 
high coupling regimes as demonstrated in Fig.~\ref{fig4} \& Fig.~\ref{fig5}. 
However, Fig.~\ref{fig6}(c) depicts that
sIPR profile maintain a low value regardless of the coupling strength reflecting a direct 
transition from the incoherent to the coherent dynamics. This observation reflects that not only the
introduction of delays but also the exact number of the delayed nodes in a network affects the 
CR and ICR distributions. As we increase the number of heterogeneously delayed nodes, the ICR expands shrinking the CR. However, for a large number of nodes having heterogeneous delays, the perturbation spreads in the entire network destroying the cohesion of the CR. This can be easily understood from Fig.~\ref{fig3} where the neighboring nodes of the delayed cluster (green shaded region) posses some undelayed and some delayed edges. This impose a dynamical ``tug-of-war" onto the neighboring the nodes. For a sufficiently large delayed node cluster, the relatively small undelayed nodes looses coherence and converts the small CR into ICR and hence chimera state lost.

\begin{figure}[t]
 \centerline{\includegraphics[width=3.8in,height=3.0in]{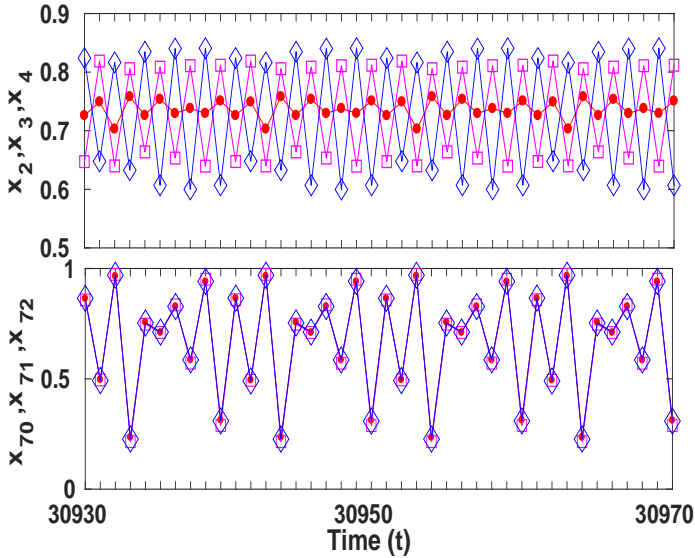}}
 \caption{(Color online) Time series of delayed and undelayed nodes of a regular ring network. A snapshot of the nodes (2,3,4,70,71,72) at particular time can be found in Fig.~\ref{fig5}(c) depicting a engineered chimera state. Time series of the node number 2,3,4 with heterogeneous delays demonstrate a non-synchronized evolution whereas bottom figure plots time series of undelayed nodes 70,71,72 depicting a synchronized evolution. Other parameters are same as in Fig.~\ref{fig5}.}
 \label{fig7}
 \end{figure}

This investigation indicates that the partial heterogeneous delays play a crucial role in the 
emergence of the chimera states. The influence of delayed nodes in the engineering of the
chimera state can be explained by comparing the time evolution of the delayed with those of
the undelayed nodes. At high coupling values, the undelayed nodes reach to a 
coherent state and the delayed nodes lag behind due to the existence of the heterogeneous delays. 
Fig.~\ref{fig7} demonstrates a typical time series of the six nodes belonging to ICR (Node 2,3, \& 4) 
and CR (Node 70,71,72), respectively, as depicted in Fig.~\ref{fig5}(c). The time series of 
undelayed nodes (node number 70,71,72) reflect a coherent synchronous evolution 
(bottom subfigure of Fig.~\ref{fig7}) whereas heterogeneously delayed nodes (node number 2,3,4) 
evolve in an incoherent fashion producing a coherent-incoherent hybrid dynamical state referred 
as chimera. The disorderly ``phase lags" introduced by the heterogeneous delays result in 
the ICR. Note that an arrangement of partial homogeneous delays will produce two clusters of nodes 
having a fixed lag between them. This in turn manifests in two CRs separated by a point discontinuity. 
To avoid such spatial states, we have considered a heterogeneous delay in our demonstration.

\section{Conclusion.}
We have provided a new approach towards control of the emergent chimera patterns in regular networks. We have
demonstrated that the location of the incoherent region coincides with the edges having the
heterogeneous delays. Furthermore, by appropriate distribution of the heterogeneous
delays, it is possible to engineer both the single-cluster and the multi-cluster chimera state. Moreover,
we have demonstrated that this control works only in the high coupling region. For both the undelayed and
the homogeneous delayed cases, the high coupling region manifests an occurrence of the coherent dynamics.
By introducing the heterogeneous delays, we can generate tailor-made chimera states in this high coupling region. Therefore, the heterogeneous delays not only causes an enhancement in the parameter regime for which chimera appears but also offers control over chimera patterns in a limited parameter range. Further, we
found that the heterogeneous delays spanned over the entire network destroy the emergence of chimera patterns for any coupling strength. The dynamical systems directly jump from the incoherent to the coherent state if all the edges of the network posses the heterogeneous delays.

To summarize, we show that in a regular network, the chimera regime can be enhanced with an
introduction of the heterogeneous delays in the edges. Importantly, the chimera patterns can be
designed by placing the delayed nodes in suitable spatial positions. Furthermore, heterogeneous
delays spanning all the edges can lead to a destruction of the chimera state in the network.

This brief report demonstrates an investigation of the impact of heterogeneously distributed delay on the emergence of the chimera states. Furthermore, it presents a method to control and design the same in a limited parameter regime. A new measure based on the spectral graph theory has been put forward for identification of the chimera states. Chimera states have recently been a topic of great interest due to their potential applicability in various systems. The investigation on chimera state is believed to be helpful in the diagnosis of several neurological disorder~\cite{brain_disease,EEG_chim}. Controlling these state is crucial to develop potential applications. This investigation sheds light on the behavior of chimera states under heterogeneous delays which are intrinsic in real-world complex systems and presents an approach to control the chimera states.

\section{Supplementary Material} \label{SM}
See supplementary material for a diagram depicting tailored chimera states for different values of $\tau_{max}$.

\section{Acknowledgment}
SJ and SG, respectively, acknowledge DST, Government of India project EMR/2016/001921 and the INSPIRE fellowship (IF150149) for financial support. 


\begin{thebibliography}{99}
\bibitem{tau} We have chosen $\tau_{max}$ to be larger than the intrinsic time scale to get a high variance in the entries of the delay matrices. As a result, a large spread in the delay values ensures that the selected delayed nodes lag considerably and randomly as compared to the undelayed nodes. Thus, in a snapshot, the selected nodes produce a high scattering owing to their different lags. However, the choice of $\tau_{max}$ does not considerably influence the results shown in the manuscript. A comparative diagram of the snapshots for different $\tau_{max}$ is presented in Sec.~\ref{SM} which demonstrate same engineered chimera state with a little different scatter pattern in the ICR region.
\end{thebibliography}
\end{document}